\documentclass[sn-mathphys-num]{sn-jnl}

\usepackage{graphicx}%
\usepackage{multirow}%
\usepackage{amsmath,amssymb,amsfonts}%
\usepackage{amsthm}%
\usepackage{mathrsfs}%
\usepackage[title]{appendix}%
\usepackage{xcolor}%
\usepackage{textcomp}%
\usepackage{manyfoot}%
\usepackage{booktabs}%
\usepackage{listings}%
\usepackage[ruled,vlined]{algorithm2e}
\theoremstyle{thmstyleone}%
\theoremstyle{thmstyletwo}%

\theoremstyle{thmstylethree}%

\begin{document}

\title[A Novel Hybrid Approach for Time Series Forecasting: Period Estimation and Climate Data Analysis Using Unsupervised Learning and Spline Interpolation]{A Novel Hybrid Approach for Time Series Forecasting: Period Estimation and Climate Data Analysis Using Unsupervised Learning and Spline Interpolation}


\author[1]{\fnm{Tanmay} \sur{Kayal}}\email{tanmay@iitmandi.ac.in}

\author*[2]{\fnm{Abhishek} \sur{Das}}\email{abhishek.das@vit.ac.in}

\author[2]{\fnm{U} \sur{Saranya}}\email{saranya.u2020@vitstudent.ac.in}

\affil[1]{\orgname{School of Mathematical \& Statistical Sciences, Indian Institute of Technology Mandi}, \orgaddress{ \city{Kamand}, \postcode{175075}, \state{Himachal Pradesh}, \country{India}}}

\affil[2]{\orgdiv{Department of Mathematics}, \orgname{School of Advanced Sciences, Vellore Institute of Technology}, \orgaddress{\city{Vellore}, \postcode{632014}, \state{Tamil Nadu}, \country{India}}}


\abstract{This article explores a novel approach to time series forecasting applied to the context of Chennai's climate data. Our methodology comprises two distinct established time series models, leveraging their strengths in handling seasonality and periods. Notably, a new algorithm is developed to compute the period of the time series using unsupervised machine learning and spline interpolation techniques. Through a meticulous ensembling process that combines these two models, we achieve optimized forecasts. This research contributes to advancing forecasting techniques and offers valuable insights into climate data analysis.}




\maketitle
\section{Introduction}
Climate forecasting plays a pivotal role in understanding and mitigating the impact of climate change on various regions worldwide. In this context, accurate prediction models are essential for anticipating climate patterns and making informed decisions to address the associated challenges. While traditional regression models have been widely employed for forecasting purposes, the inherent dynamic and temporal nature of climate data necessitates a more nuanced approach. Time series analysis, a powerful statistical technique designed to analyze sequential data points, emerges as a compelling alternative for climate forecasting.
Unlike regression models that assume independence among data points, time series analysis acknowledges the temporal structure inherent in climate data. This temporal dependency is vital for capturing the cyclical, seasonal, and trend-related patterns that characterize climate phenomena.

\noindent
Chennai, a coastal metropolis in southern India, experiences a diverse range of climatic conditions, ranging from tropical monsoons to intense heatwaves. The dynamic and multifaceted nature of Chennai's climate demands forecasting models capable of capturing intricate temporal dependencies. In the literature, one can find many models introduced to address a varieties of time series data.

\noindent
In \cite{r1}, the authors employed forecasting methods based on four model classes: autoregressions (with and without unit root pretests), exponential smoothing, artificial neural networks, and smooth transition autoregressions. Autoregression with unit root pretests demonstrated the best overall performance, and combining it with other models further improved predictive accuracy. The paper \cite{r2} introduces the Quarterly Earnings Prediction (QEP) model, a new univariate model based on epsilon support vector regression. The QEP model exhibits significant performance under various conditions.

\noindent
In \cite{r3}, the objective is to build a framework for forecasting univariate time series and conduct a performance comparison between different models.
The study in \cite{r4} examines trend and stationarity, identifying autoregressive integrated moving average (ARIMA) models as suitable after finding randomness and non-stationarity within the time series.
Authors in \cite{r5} compare seasonal autoregressive integrated moving average (SARIMA) and Holt-Winter's Exponential Smoothing approaches for highly accurate consumer transaction forecasts in Store X, with SARIMA demonstrating continued accuracy.

\noindent
In \cite{r6}, the focus is on forecasting tuberculosis (TB) incidence in Qinghai province by combining a  SARIMA with a neural network nonlinear autoregression (NNNAR) for enhanced detection.

\noindent
Using the Seasonal Autoregressive Moving Average model, \cite{r7} predicts the future behavior of monthly average rainfall and temperature in South Asian countries, employing the Mann-Kendall trend test and Sen's Slope Estimator.
The study in \cite{r8} uses univariate time series analysis, including SARIMA and exponential smoothing (ES) models, to forecast accident rates in Pakistan. ES is found to better suit accident data.

\noindent
In \cite{r9}, a Holt-Winters Exponential Smoothing strategy is proposed to forecast economic time series with significant trends and seasonal patterns, showing promise in capturing different patterns in the data.

\noindent
The author in \cite{r10} utilizes the Holt-Winters exponential smoothing additive model to predict seasonal time series data for the number of passengers departing from Hasanudin Airport, demonstrating reduced forecasting errors by incorporating trend and seasonal patterns.

\noindent
Collectively, these studies showcase the diverse applications of univariate forecasting models across various domains, highlighting the importance of choosing appropriate models based on the characteristics of the data and the specific requirements of the forecasting task.

\noindent
In this research article, we delve into the development and application of a new time series forecasting method, which is applied on  Chennai climate data. To mention briefly, our proposed method consists of two different established time series models using seasonality and period. It is worthwhile to mention that the algorithm we develop to compute the period is a combination of unsupervised machine learning and spline interpolation technique. Further, we construct an ensembling method considering these two models to obtain the optimized forecasts.

\noindent
The paper is arranged as follows:
Section \ref{S2} consists of necessary theoretical backgrounds of the proposed method.
In section \ref{S3}, we discuss the problem statement along with the detailed approach.
It also includes the analysis of the results obtained through the proposed method. Finally, the paper ends with a conclusion, in Section \ref{S4}.

\section{Theoretical Background }\label{S2}
Here, we introduce few terminologies which are very frequent in the theory of time series forecast and used to describe our proposed model.



\noindent
\textbf{Auto-regressive:}
The Auto-Regressive model predicts future values based on past values, incorporating error or random noise: $x_t=c+\phi_1x_{t-1}+\cdots+\phi_px_{t-p}+\varepsilon_t$.\\

\noindent
\textbf{Moving average:}
The Moving Average is calculated by averaging values within a sliding window: $x_t=\varepsilon_t+\theta_1\varepsilon_{t-1}+\cdots+\theta_q\varepsilon_{t-q}$.\\

\noindent
\textbf{Z-score method:}
The Z-score ($z$) of an observation ($X$) is calculated as $z=(X-\mu)/\sigma$.\\

\noindent
\textbf{Linear Interpolation:}
Linear interpolation is employed to treat outliers by fitting linear polynomials, generating new data points using the formula $y = y_1 + ( x - x_1 ) ( y_2 - y_1 ) / ( x_2 - x_1)$.\\

\noindent
\textbf{SARIMA model:}
SARIMA model is introduced to handle seasonality in the data. In this article, SARIMA model is denoted as,
SARIMA $(p,\,d,\,q,\,tr)\,(P,\,D,\,Q)\,m$,\\

\noindent
where,
\begin{itemize}
    \item  $p$ = Number of trend AR terms, $P$ = Number of seasonal AR terms,
     \item $d$ = Number of trend differences, $D$ = Number of seasonal differences,
     \item $q$ = Number of trend MA terms, $Q$ = Number of seasonal MA terms,
    \item  $m$ = Length of the season 
    \item  $tr$ is a variable that contains four types of different parameter values `$n$', `$c$', `$t$' and `$ct$', where `$n$', `$c$', `$t$' and `$ct$' denote `no trend', `constant trend', `linear trend' and `constant and linear trend', respectively.
\end{itemize}
    
\noindent
An algorithmic approach (\ref{alg:sarima-part1}), is used to determine the parameters for SARIMA model.\\


\begin{algorithm}
\caption{SARIMA Parameter Optimization Algorithm }
\label{alg:sarima-part1}
\SetKwInput{Input}{Input}
\SetKwInput{Output}{Output}
\textbf{Step 1:} Take the inputs for training and validation data\;
\textbf{Step 2:} Choose values for the coefficients for AICc ($C_1$) and error metric ($C_2$) with the constraints $C_j\in [0,\,1]$ and $\sum\limits_{j=1}^k C_j=1$, where $j=1,\,2$\;
\textbf{Step 3:} Set ranges for $p,\,d,\,q,\,P,\,D,\,Q$ and set value for `$m$'\;
\textbf{Step 4:} Set $tr$=`$n$',\,`$c$',\,`$t$',\,`$ct$' \;
\textbf{Step 5:} Compute all possible combinations for the parameters from steps 3-4 and store them in a variable (say $V$) \;
\ForEach {$l$ in $V$} {
    \textbf{Step 6:} Fit a SARIMA model\;
    \textbf{Step 7:} Calculate the model's AICc (say $v_1$)\;
    \textbf{Step 8:} Perform a forecast for the validation set\;
    \textbf{Step 9:} Compute SMAPE between the forecast and the actual values in the validation set, say the error is $v_2$\;
    \textbf{Step 10:} Calculate the Combined Effect using $C_1v_1+C_2v_2$ and store for each iteration\;
}
\textbf{Step 11:} Pick the parameter values having minimum combined effect.
\end{algorithm}

\noindent
\textbf{Holt-Winter Exponential Smoothing (HWES) method:} This approach comes in two variations, each with a distinct treatment of the seasonal component. The additive method is recommended when seasonal fluctuations remain relatively constant across the entire series, while the multiplicative method is preferred when these fluctuations change in proportion to the level of the series.

\noindent
The component form for the additive method is as follows:
\begin{eqnarray}\label{hwes1}
  \hat{y}_{t+h|t} &=& \ell_{t} + hb_{t} + s_{t+h-m(k+1)} \\\nonumber
  \ell_{t} &=& \alpha(y_{t} - s_{t-m}) + (1 - \alpha)(\ell_{t-1} + b_{t-1})\\\nonumber
  b_{t} &=& \beta^*(\ell_{t} - \ell_{t-1}) + (1 - \beta^*)b_{t-1}\\\nonumber
  s_{t} &=& \gamma (y_{t}-\ell_{t-1}-b_{t-1}) + (1-\gamma)s_{t-m},
\end{eqnarray}

\noindent
where $k$ ensures that the estimates of the seasonal indices used for forecasting come from the final year of the sample by being the integer component of $(h-1)/m$.

\noindent
The component form for the multiplicative method is as follows:
\begin{eqnarray}\label{hwes2}
  \hat{y}_{t+h|t} &=& (\ell_{t} + hb_{t})s_{t+h-m(k+1)} \\ \nonumber
  \ell_{t} &=& \alpha \frac{y_{t}}{s_{t-m}} + (1 - \alpha)(\ell_{t-1} + b_{t-1})\\\nonumber
  b_{t} &=& \beta^*(\ell_{t}-\ell_{t-1}) + (1 - \beta^*)b_{t-1}                \\\nonumber
  s_{t} &=& \gamma \frac{y_{t}}{(\ell_{t-1} + b_{t-1})} + (1 - \gamma)s_{t-m}.
\end{eqnarray}
The forecast equations (\ref{hwes1})-(\ref{hwes2}), together with three smoothing equations for the level $l_t$, trend $b_t$, and seasonal component $s_t$, are all included in the HWES technique. In the equations, smoothing parameters are $\alpha,\,\beta^*$ and $\gamma$. We use the symbol $m$ to represent the number of seasons in a year. One can refer \cite{r14,r15} for more details.

\noindent
Here, in algorithm \ref{alg:sarima-part2}, we discuss the parameter optimization process for HWES method.

\begin{algorithm}
\caption{HWES Parameter Optimization Algorithm }
\label{alg:sarima-part2}
\SetKwInput{Input}{Input}
\SetKwInput{Output}{Output}
\textbf{Step 1:} Take the inputs for training and validation data\;
\textbf{Step 2:} Choose values for the coefficients for AICc ($C_1'$) and error metric ($C_2'$) with the constraints $C_j'\in [0,\,1]$ and $\sum\limits_{j=1}^k C_j'=1$, where $j=1,\,2$\;
\textbf{Step 3:} Set ranges for smoothing level, smoothing slope, smoothing seasonal and set value for `$m$'\;
\textbf{Step 4:} Set trend = [`add', `mul'] and seasonal = [`add', `mul'], where `add' and `mul' are additive and multiplicative trends, respectively\;
\textbf{Step 5:} Compute all possible combinations for the parameters from steps 3-4 and store them in a variable (say $V$) \;
\ForEach {$l$ in $V$} {
    \textbf{Step 6:} Fit a HWES model\;
    \textbf{Step 7:} Calculate the model's AICc (say $v_1'$)\;
    \textbf{Step 8:} Perform a forecast for the validation set\;
    \textbf{Step 9:} Compute SMAPE between the forecast and the actual values in the validation set, say the error is $v_2'$\;
    \textbf{Step 10:} Calculate the Combined Effect using $C_1'v_1'+C_2'v_2'$ and store for each iteration\;
}
\textbf{Step 11:} Pick the parameter values having minimum combined effect.
\end{algorithm}

\noindent
\textbf{Kalman filter:} The Kalman filter is an optimal recursive algorithm for estimating the state of a linear dynamic system from a series of noisy measurements. It is widely used in various fields, including control systems, navigation, finance, and signal processing.\\

\noindent
{\it Prediction Step:}
\[
\hat{x}_k^- = F_k \hat{x}_{k-1} + B_k u_k
\]
\[
P_k^- = F_k P_{k-1} F_k^T + Q_k
\]

\noindent
Where:
\begin{itemize}
  \item $\hat{x}_k^-$ is the predicted state.
  \item $F_k$ is the state transition matrix.
  \item $B_k$ is the control-input matrix.
  \item $u_k$ is the control input at time $k$.
  \item $P_k^-$ is the predicted error covariance.
  \item $Q_k$ is the process noise covariance.
\end{itemize}

\noindent
{\it Update Step:}
\[
K_k = P_k^- H_k^T (H_k P_k^- H_k^T + R_k)^{-1}
\]
\[
\hat{x}_k = \hat{x}_k^- + K_k(z_k - H_k \hat{x}_k^-)
\]
\[
P_k = (I - K_k H_k) P_k^-
\]

\noindent
Where:
\begin{itemize}
  \item $K_k$ is the Kalman gain.
  \item $H_k$ is the measurement matrix.
  \item $R_k$ is the measurement noise covariance.
  \item $z_k$ is the measurement at time $k$.
  \item $\hat{x}_k$ is the updated state estimate.
  \item $P_k$ is the updated error covariance.
\end{itemize}
The Kalman filter minimizes the mean squared error of the estimated state, making it optimal for linear systems with Gaussian noise. For more detail, refer \cite{r11,r12}.\\

\noindent
\textbf{$K$-means clustering:} This algorithm aims to partition a dataset into $K$ pre-defined clusters, where each data point belongs to only one cluster based on its similarity to other points \cite{r13}. This technique is particularly useful for exploring unlabeled data, like Chennai climate records, where inherent structures might be unknown. The steps for this algorithm are as follows:
\begin{algorithm}
\caption{$K$-means clustering }
\label{alg:kmeans-part1}
\SetKwInput{Input}{Input}
\SetKwInput{Output}{Output}
\textbf{Step 1:} Select $K$ points as initial centroids.\;
\textbf{Step 2:}

{\bf repeat}

Form $K$ clusters by assigning each point to its closest centroid.

Recompute the centroid of each cluster.

{\bf until} Centroids do not change.
\end{algorithm}


\noindent
\textbf{Exploratory data analysis (EDA):}
EDA is a crucial step involving the analysis of data characteristics through visualization and summary. Statistical measures such as mean and standard deviation are employed in this phase.\\

\noindent
\textbf{Data splitting:}
For effective machine learning, a dataset is partitioned into training, test, and validation sets. The training set is employed to train the model and establish its key parameters, while the test set evaluates the model's generalizability. This involves assessing the model's ability to recognize patterns in previously unseen data.\\

\noindent
\textbf{Data Pre-processing and cleaning:}
In the initial phase of data mining, data pre-processing involves transforming raw data into an analyzable format. In our considered dataset, columns denoting day, month, and year were consolidated into a single column named `date'. The pre-processed data is then selected for further analysis.

\noindent
In the data cleaning stage, the dataset undergoes scrutiny for null or missing values. Specifically, the `date' column is examined to identify any missing or duplicate dates. Additionally, the dataset is scanned for outliers, which are addressed through linear interpolation based on the Z-score method.\\

\noindent
\textbf{Outlier analysis:}
Outliers, inaccurate observations in the data, are detected through box and scatter plots. The Z-score method is employed for outlier identification.\\

\noindent
\textbf{Symmetric mean absolute percentage error (SMAPE):}
The SMAPE is used also to measure the accuracy of forecast. SMAPE is the most commonly used error metrics and it is the measure of relative errors.\\

$$SMAPE=\dfrac{1}{n}\times \displaystyle\sum \dfrac{|actual\,value-forecast\,value|}{(|actual\,value|+|forecast\,value|)/2}. $$

\noindent
\textbf{ Root mean square error (RMSE):} RMSE is the standard way to measure the error of a model in time series.

 $$RMSE=\sqrt{\sum_{i=1}^n\dfrac{(\hat{y}_i-y_i)^2}{n}}.$$\\

 \noindent
\textbf{Cubic B-spline interpolation:}
This is defined by a set of control points $\{P_i\}_{i=0}^n$, where $P_i = (x_i, y_i)$ in 2D space. The B-spline curve $C(t)$ is given by the formula:
\[
C(t) = \sum_{i=0}^{n} N_{i,3}(t) \cdot P_i
\]
where $N_{i,3}(t)$ are the cubic B-spline basis functions. These basis functions are defined recursively as:
\[
N_{i,1}(t) =
\begin{cases}
1 & \text{if } t_i \leq t < t_{i+1} \\
0 & \text{otherwise}
\end{cases}
\]
\[
N_{i,k}(t) = \frac{t - t_i}{t_{i+k-1} - t_i} \cdot N_{i,k-1}(t) + \frac{t_{i+k} - t}{t_{i+k} - t_{i+1}} \cdot N_{i+1,k-1}(t)
\]
for $k > 1$, and where $t_0, t_1, \ldots, t_{n+3}$ are the knot vector points. These knot vector points are chosen such that $t_i \leq t_{i+1}$ for all $i$, and $t_{n+3} - t_0$ is the total range of parameter $t$ \cite{r16}.
\section{Proposed Model}\label{S3}
\begin{figure}[h]
    \centering
\includegraphics[width=0.75\textwidth]{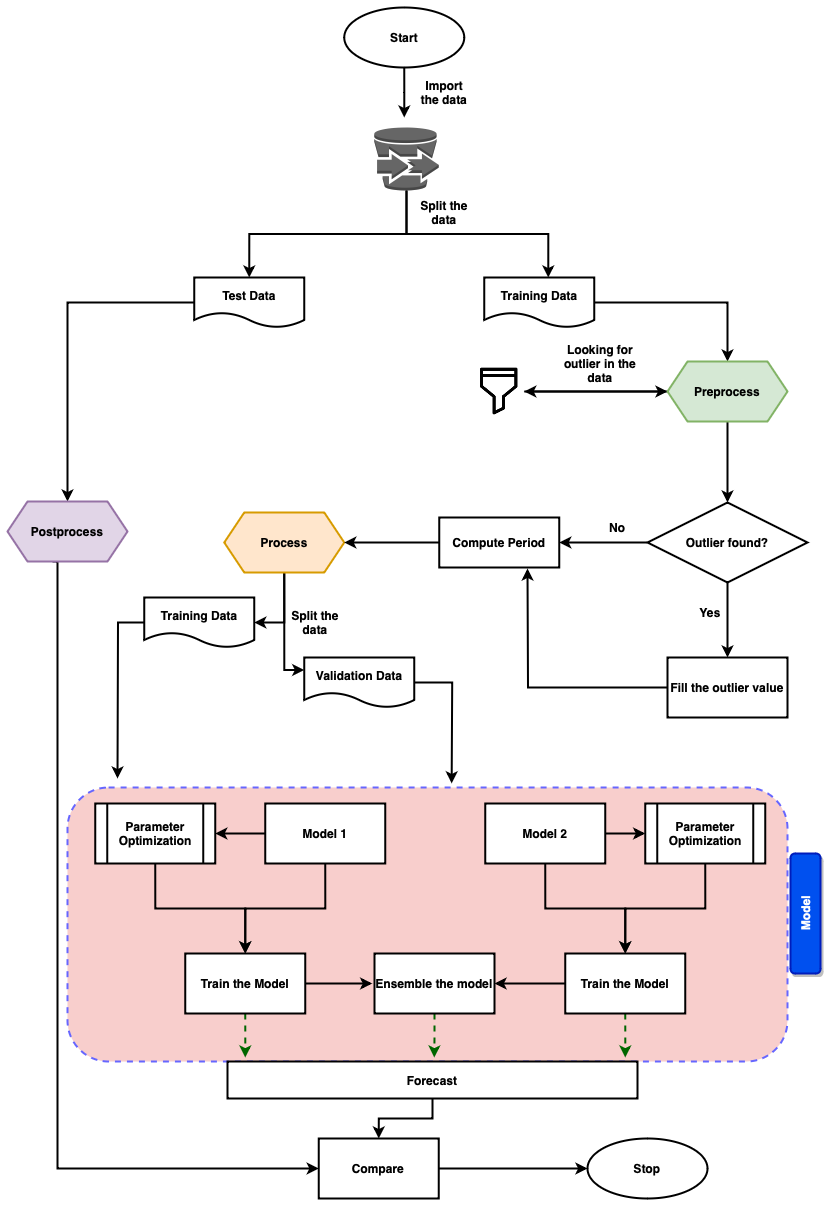}
    \caption{Flow for the proposed approach}
    \label{fig_flow}
\end{figure}

To begin this section, one can see the Figure \ref{fig_flow} to get an overall glimpse of the proposed approach.
Now, we discuss the proposed approach elaborately.
\subsection{Exploratory data analysis}
We initiate our study by extracting daily temperature data from January $1$, $2022$, to January $9$, $2023$, from the Visual Crossing Weather API \footnote{Data Source: Visual Crossing Weather API: https://www.visualcrossing.com/weather-api}. To ensure a robust model, we perform a comprehensive EDA and preprocessing of the dataset. For training purposes, we consider the data up to December $30$, $2022$.\\

\begin{figure}[h]
    \centering
    \includegraphics[width=0.5\textwidth]{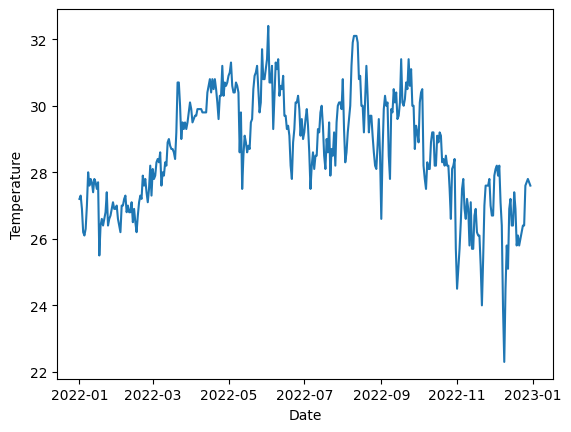}
    \caption{Visualization of the training data}
    \label{fig1}
\end{figure}

 \noindent
\textbf{Data Summary:}
Figure \ref{fig1} displays the training data, which exhibits a mean and standard deviation of 28.64 and 1.69, respectively.

\begin{figure}[h]
    \centering
    \includegraphics[width=0.5\linewidth]{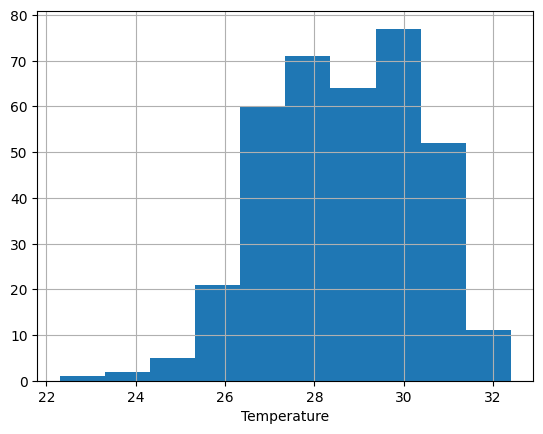}
    \caption{Histogram plot}
    \label{fig3}
\end{figure}

 \noindent
The histogram in Figure \ref{fig3} provides insights into the distribution of the training data. \\


\begin{figure}[hbt!]
  \centering
  \includegraphics[width=0.8\textwidth]{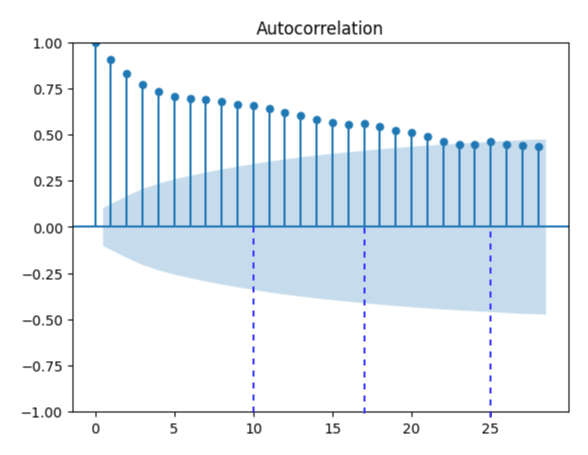} 
  \caption{Autocorrelation plot for the training data}
  \label{fig:acf}
\end{figure}

\begin{figure}[h]
    \centering
    \includegraphics[width=0.5\linewidth]{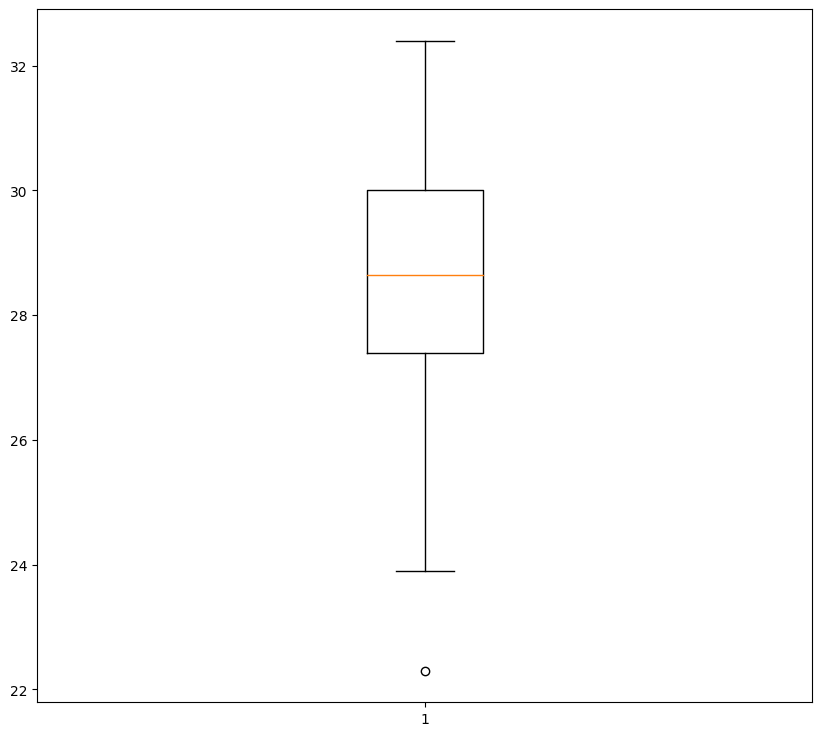}
    \caption{Box plot for outlier detection}
    \label{fig4}
\end{figure}

\begin{figure}[h]
    \centering
    \includegraphics[width=0.5\linewidth]{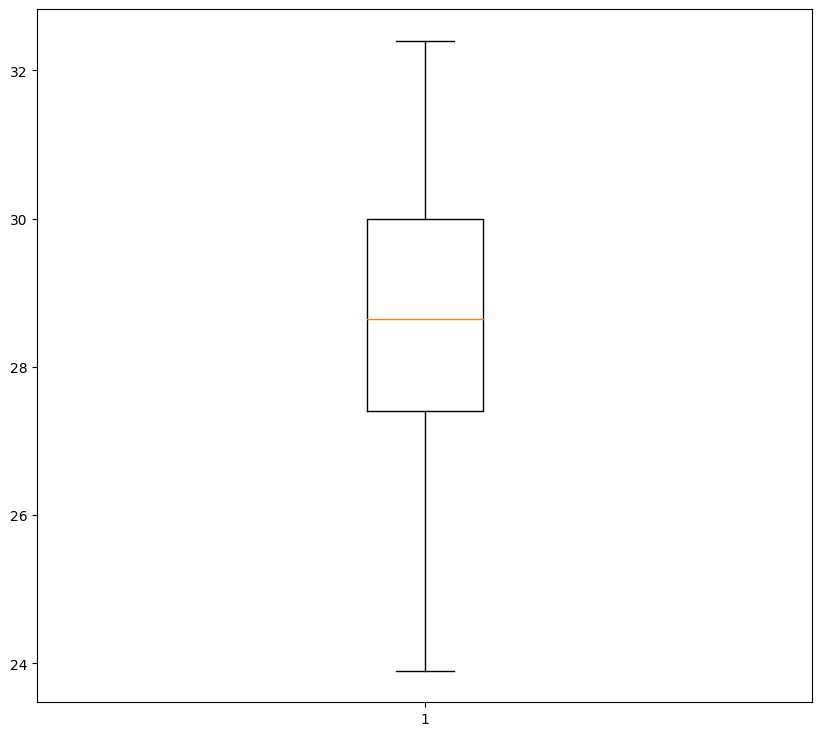}
    \caption{Box plot after outlier adjustment}
    \label{fig5}
\end{figure}

 \noindent
\textbf{Outlier handling:}
 We detect an anomaly present on 2022-12-09, with a temperature of 22.3 using Z-statistic method. One can visualize the presence of outlier in the data through Figure \ref{fig4}. Employing polynomial interpolation with order = 1, we fill the outlier value, resulting in an interpolated temperature of 24.2 (Figure \ref{fig5}). Furthermore, after interpolating the outlier value, we observe from the Figure~\ref{fig:acf} that the possible periods for the data are 10, 17 and 25 days, respectively.


\noindent
\textbf{Data Transformation:}
We transformed the training data through a series of steps:\\

\begin{enumerate}
     \item Log transformation with base 10.
    \item  Differencing with order 1, \emph{i.e.,} ${Y'}_{t_1}=Y_{t_2}-Y_{t_1},\,{Y'}_{t_2}=Y_{t_3}-Y_{t_2}$, and so on,  to achieve stationarity.
    \item Splitting the transformed data into Training Data (top 95\%) and Validation Data (last 5\%).
\end{enumerate}
   
\begin{figure}[h]
    \centering
    \includegraphics[width=0.5\linewidth]{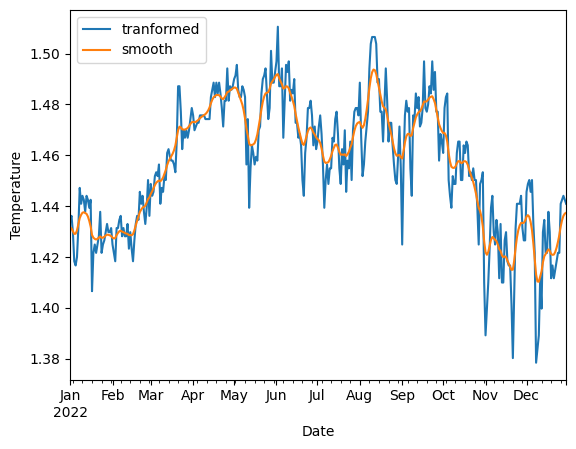}
    \caption{Kalman Filter application on the data}
    \label{fig6}
\end{figure}

\noindent
\textbf{Augmented Dickey-Fuller (ADF) Test and Period Computation:}
To assess the stationarity of the data, we conduct an ADF test, necessitating knowledge of the data period. Applying Kalman Filter smoothing with transition matrix = 1 and covariance = 0.1, as shown in Figure \ref{fig6}, helped compute the period. For this purpose, we use a Python library called `\texttt{pykalman}'.\\

\begin{figure}[h]
    \centering
    \includegraphics[width=0.5\linewidth]{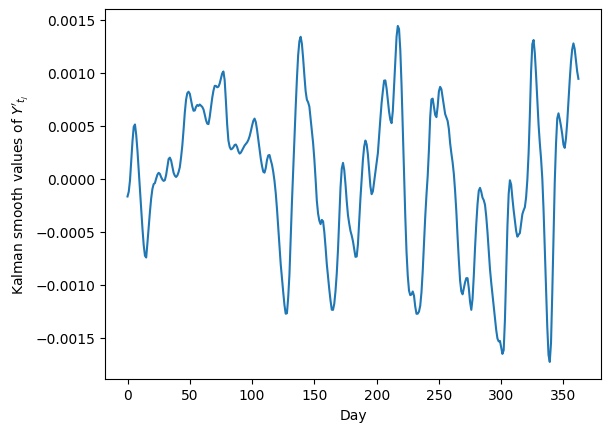}
    \caption{Visualisation of the processed data after percentage change}
    \label{fig7}
\end{figure}

\noindent
Next, we take the percentage change \emph{i.e.},${Y'}_{t_1}=\frac{Y_{t_2}}{Y_{t_1}}-1,\,{Y'}_{t_2}=\frac{Y_{t_3}}{Y_{t_2}}-1$, and so on, which reflects in the Figure \ref{fig7}.

\begin{figure}
    \centering
    \includegraphics[width=0.5\linewidth]{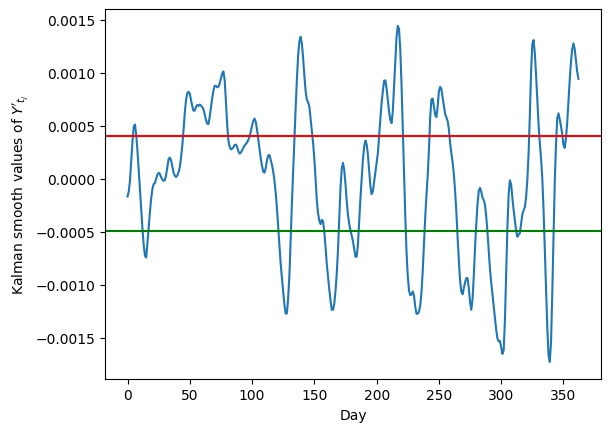}
    \caption{Clustering with $K$-means algorithm}
    \label{fig8}
\end{figure}

\noindent
\textbf{$K$-means clustering and day difference:}
In Figure \ref{fig8}, we illustrate the use of the $K$-means algorithm with \( K = 3 \) to cluster the given data points into three distinct groups. $K$-means is a popular unsupervised learning algorithm that partitions data into clusters based on their similarities. The algorithm iteratively assigns data points to one of the \( K \) clusters, minimizing the distance between each data point and the centroid of its assigned cluster. In our case, we use \( K = 3 \) to divide the data into three clusters: lower, middle, and higher, based on the characteristics of the data points.

\noindent
Once the data points were clustered, we focused on analyzing the lower and higher clusters separately. These clusters represent different segments of the data, which allow us to examine the trends more clearly. The higher cluster corresponds to data points that exhibit higher values, while the lower cluster represents those with lower values. By isolating these clusters, we perform more targeted analyses on each group's behavior.

\begin{figure}
    \centering
    \includegraphics[width=0.5\linewidth]{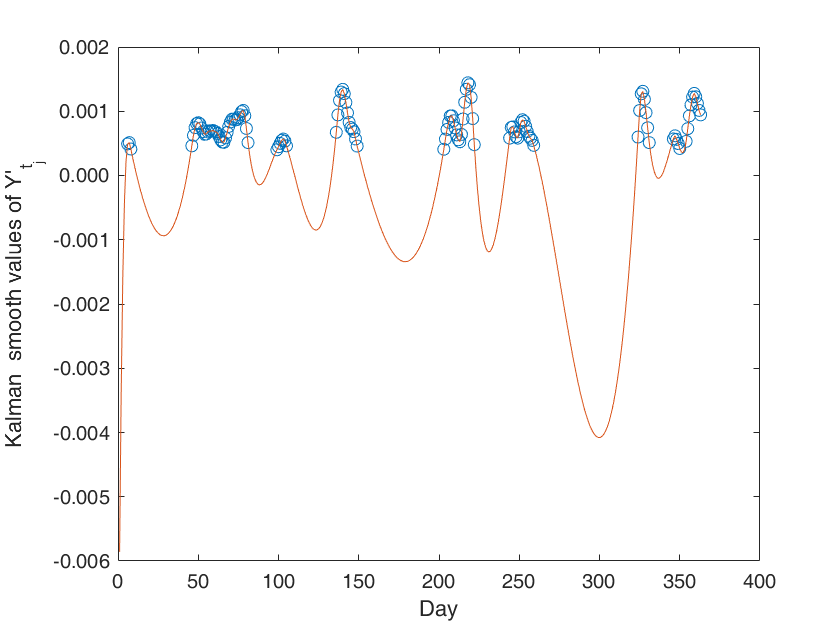}
    \caption{Upper cluster polynomial}
    \label{fig10}
\end{figure}

\noindent
To further analyze the higher cluster points, we employ a cubic B-spline interpolation technique which is discussed in Section \ref{S2}. This method constructs smooth, piecewise polynomial functions that pass through or near the higher cluster points, as shown in Figure \ref{fig10}. The advantage of using B-splines is that they provide a smooth approximation of the data and are particularly effective for identifying critical points, such as local maxima. By applying this technique, we are able to accurately locate the local maxima in the higher cluster data, which are essential for understanding periodic trends in the dataset.

\begin{figure}
    \centering
    \includegraphics[width=0.5\linewidth]{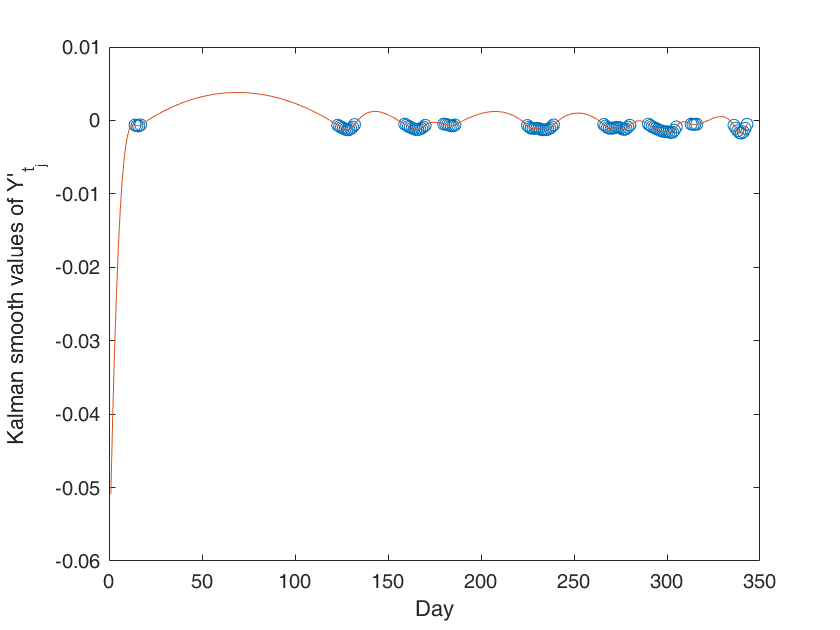}
    \caption{Lower cluster polynomial}
    \label{fig11}
\end{figure}

\noindent
Similarly, we use the same cubic B-spline approach to analyze the lower cluster, but in this case, we focus on identifying the local minima. Figure \ref{fig11} shows the polynomial corresponding to the lower cluster, which allows us to pinpoint the local minima points. These maxima and minima provide important insights into the data's periodic behavior and help us better understand fluctuations over time.


\begin{figure}
    \centering
    \includegraphics[width=\linewidth]{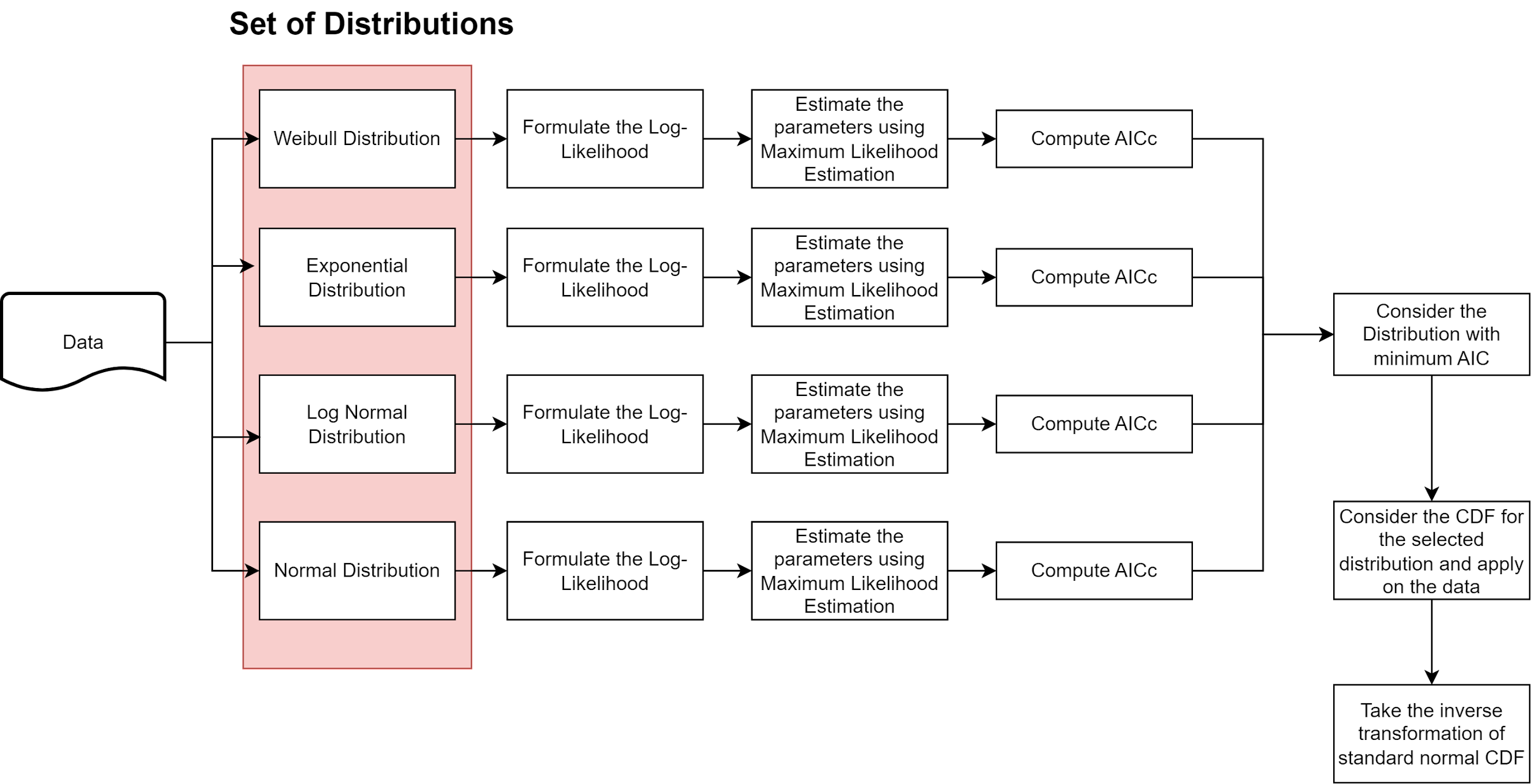}
    \caption{Transformation technique into standard normal}
    \label{fig9}
\end{figure}

\noindent
Next, we compute the day differences and transform that to a standard normal distribution. A graphical approach is given in Figure \ref{fig9}.\\

\noindent
\textbf{Exponential Distribution:} The cumulative distribution function we use for the analysis is known as exponential distribution, which is defined as $F(x)=1-\exp{(-x/\lambda)}$, where $x,\,\lambda>0$. Through rigorous analysis, we observe a preferable fit of the Exponential distribution to the dataset. Constructing a 95\% confidence interval, we convert the range back into the original distribution, obtaining a mean period of 23 days.

\noindent
To better understand the periodic patterns within the 23-day period, we present the decomposition plot. This plot provides insights into the trend, seasonality, and residual components over the specified period.

\begin{figure}[hbt!]
  \centering
  \includegraphics[width=0.8\textwidth]{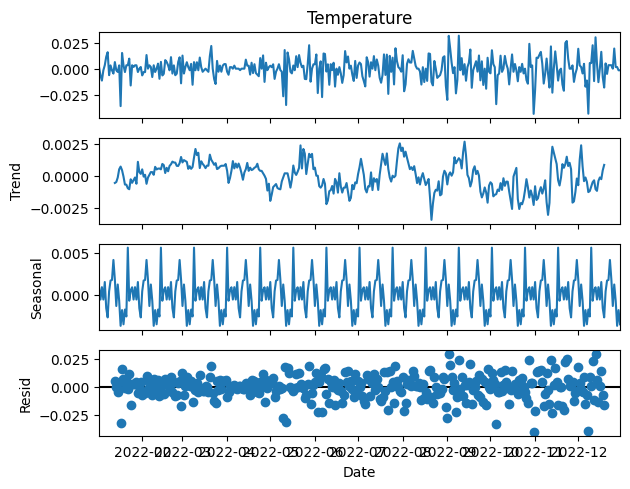} 
  \caption{Decomposition plot for the 23-day period, where `temp' denotes the transformed temperature data }
  \label{fig:decomposition_plot}
\end{figure}

\noindent
Figure~\ref{fig:decomposition_plot} visually represents the decomposition of the time series data, offering a clear depiction of the underlying patterns. This decomposition aids in understanding the cyclic behavior within the 23-day period, facilitating more informed forecasting decisions.

\subsection{Model evaluation}
Here, we give the quantitative analysis of the model.
Following the ADF test, it was determined that the raw training data lacked stationarity (p-value = $0.488$). However, after the prescribed transformations, the data exhibits the stationarity with a period of $23$ days. To further enhance predictive capabilities, we apply the algorithm for SARIMA model with the range for $p=q=P=Q \in [0,2]$, $d=0$ as differencing is done once, $D \in [0,1]$, $C_1=0.6$ and $C_2=0.4$. Finally, the obtained values for the parameters are  $p=2, ~q=1, ~tr=\text{`n'}, P=0, ~Q=1$, ~ $d=0$ and $D=0$.

\noindent
The second model employed was the HWES model with the parameters' ranges, smoothing level = smoothing slope = smoothing seasonal $\in [0, 0.8]$ with the step size $0.1$.

\noindent
Finally, we obtain the parameters: smoothing level = $0.0$, smoothing slope = $0.2$, smoothing seasonal=$0.0$, both trend and seasonal are `add' and
$C_1=0.8$ and $C_2=0.2$.

\noindent
Now, we consider the fitted values from both the SARIMA (as model 1) and HWES (as model 2), converting them back to the original format. Subsequently, a linear regression model was fit, denoted as:

\[ Y_t = \hat{\beta_0} + \hat{\beta_1}X_{t1} + \hat{\beta_2}X_{t2} + \epsilon_t. \]

\noindent
Here, $ Y_t $ is the actual training data, and $ X_{t1} $ and $ X_{t2} $ are the fitted values (in original form) from the SARIMA and HWES, respectively. After performing the Least Squares Estimation (LSE), the estimates $\hat{\beta_0}$, $\hat{\beta_1}$, and $\hat{\beta_2}$ are obtained as $38.34$,\,$-0.66$ and $0.24$, respectively, where the results are rounded up to second digits. The ensemble of the two models is thus achieved.

\noindent
In Table \ref{table1} the error estimates on the test data for the ensemble of the SARIMA and HWES models are assessed using the SMAPE and RMSE.
\begin{table}[h!]
\centering
\begin{tabular}{|l|r|r|}
\hline Model & SMAPE & RMSE \\
\hline SARIMA & 3.16 & 1.04 \\
\hline HWES & 2.71 & 0.96 \\
\hline Ensemble & 2.29 & 0.74 \\
\hline
\end{tabular}
\caption{Error metrics for three different models}
\label{table1}
\end{table}

\noindent
These results signify the effectiveness of the proposed ensemble univariate forecasting method, showcasing improved predictive accuracy compared to individual models. The combination of SARIMA and HWES contributes to a robust forecasting framework for Chennai climate data.

\section{Conclusion and Future Work}\label{S4}
In this study, we introduced and applied a novel time series forecasting method on Chennai climate data. Our approach, a combination of two well-established time series models addressing seasonality and periods, showcased promising results in predicting temperature trends. Notably, the algorithm devised to compute periods, integrating unsupervised machine learning and spline interpolation, contributed to the accuracy of our forecasts. The ensembling technique, merging the strengths of SARIMA and HWES, further enhanced predictive performance. The low SMAPE and RMSE values for the ensemble model underscored its effectiveness in capturing the intricate patterns present in Chennai's climate data.

\noindent
As a future work, one can use the proposed forecasting method for a longer time span or different geographic locations, which can provide a broader perspective on its adaptability. The consideration of external factors, such as meteorological events or urban development, might also contribute to refining the forecasting model.

\noindent
In conclusion, the presented method lays the foundation for future advancements in univariate forecasting for climate data. The iterative refinement of algorithms and models, coupled with the exploration of diverse datasets, will undoubtedly contribute to the continual improvement and applicability of forecasting techniques in the field of climate science.

\section*{Declarations}
\begin{itemize}
\item Funding: Not available
\item Competing interests: No conflict of interest.
\item Ethics approval: Not applicable, as this work did not involve humans or animals.
\item Data availability: Information related to Data, analysed during this study are included in the supplementary material.
\item Author contribution: Tanmay Kayal:
Conceptualization, Data curation, Formal analysis, Methodology, Validation,
Abhishek Das: Conceptualization, Formal analysis, Project administration, Supervision,     Saranya U.: Methodology, Resources, Initial draft. All authors reviewed the manuscript.
\end{itemize}

\end{document}